# Threshold Quantum Cryptography

Subhash Kak

**Abstract:** Most current research on quantum cryptography requires transmission and reception of single photons that creates severe implementation challenges and limits range. This paper argues for the development of *threshold quantum cryptography* protocols in which the system is secure so long as the number of photons being exchanged between Alice and Bob is below a specified threshold. We speak of a (p-k-n) threshold system where if the number of photons exchanged is less than p, the system is completely secure, when it is between p and k, the system is partially secure, and when it exceeds k, the system is insecure. The BB84 protocol is (1-1-1) whereas the three-stage protocol appears to be (p-4p-n), where p is the least number of photons necessary to determine the polarization state of identically prepared photons. New quantum cryptography systems should be sought that provide greater flexibility in the choice of p and k.

## INTODUCTION

The BB84 protocol of quantum cryptography is based on transmission of single photons carrying one-bit of information defined with respect to two sets of orthogonal bases [1]. This creates many engineering challenges and limits the applicability of quantum cryptography. Due to the Heisenberg Uncertainty Relations, the energy of photons, or equivalently their number, and the times at which they are emitted cannot be simultaneously fixed. To produce photons that appear singly in the specified time window with high probability requires the use of extremely low power lasers which, in turn, brings down the effective clock rate to just a few thousand bits per second making BB84 useless as a method of *quantum communication* (for which the modem bit rates can be hundreds of millions of bits per second). Furthermore, BB84 type systems require single-photon detectors that can be blinded to hack the system [2],[3].

Due to the low rates, BB84 and its variants are used only for key exchange and limited to distance of 100 kms or so. Once the key has been exchanged between the two parties, the actual encryption is performed using some classical algorithm. But this means that the advantage of the quantum regime is no longer central to security as the classical algorithm can be broken by employing sufficient computational power. In other words, BB84 based key-exchange provides no advantage over classical public key cryptography algorithms. The gap between mathematical theory and non-ideal components that must be used in the implementation was summarized thus recently [4]: "the field, similar to non-quantum modern cryptography, is going to split in two directions: those who pursue practical devices may have to moderate their security claims; those who pursue ultimate security may have to suspend their claims of usefulness."

The idea of using fiber for single photons can be critiqued also from the perspective of information theory. To transmit single photons is to admit two givens: noise is absent, i.e. the SNR is infinite, and that the medium is completely dark and, by implication, fully protected. In contrast, modern communications practice is based on the premise of real-world noisy systems that are unprotected. (If the fiber is fully protected, one should not need any cryptographic transformation for the security needs can be met by incorporating strong intrusion detection.) Current quantum cryptography implementations go against the philosophy and spirit of communications engineering practice.

In view of this, I propose that new quantum cryptography protocols be sought that provide threshold security based on practical considerations. Quantum cryptography should allow the use of multiple photons as long as their number is below a specified value, say p. The actual number to be used should be such so that it leaves open the possibility of loss of photons in transmission.

## THRESHOLD SYSTEMS IN CLASSICAL CRYPTOGRAPHY

The idea of threshold systems is an old concept in classical cryptography where it shows up in different guises. The most familiar example is the security protocol required to open a bank vault. This necessitates the usage of keys in the same session by several designated officers. In opening a safe deposit box, two keys are needed: the bank officer's and the box-owner's, and the box-owner may additionally have to show a suitable identification card. A general threshold cryptography system may specify the order in which keys are to be applied, the number of keys to be used, and the minimum number necessary to open the box.

The k out of n threshold system is popular [5],[6]. In it the signal is divided up in n parts so that if k of these are brought together, the signal can be reconstructed whereas if fewer than k parts are available, no information about the signal leaks through. Another scheme is the p-k-n system in which the data is divided into n pieces so that fewer than p pieces do not reveal any information, p or more than p but less than k pieces reveal some information and at least k pieces are required to reveal the entire data (n-k pieces are redundant, which are introduced to make the system robust and fault-tolerant). In a true threshold system the values of k and n are different.

The k out of n scheme is equivalent to a k-k-n scheme. A 1-1-n distribution scheme creates n replicas of the original data and retrieving one of the replicas provides the original data. A 1-n-n scheme divides the data into pieces of the size of the original data while a n-n-n scheme creates n pieces each of the same size as that of the original data but all the n pieces are required for data reconstruction; early schemes included 1-k-n [7] as well as more general values [8]. Threshold schemes introduce and exploit redundancy but doing so reduces efficiency. There are newer schemes that use the redundant bits to store additional information thus enhancing efficiency [9]-[11].

In some threshold schemes an additional layer of security may be provided by using an encryption key to encrypt the data before dividing it into pieces. The encryption key is then stored using a threshold secret sharing scheme of type k-k-n and the encrypted data is divided into n pieces using 1-k-n information dispersal scheme. Such a hybrid scheme balances concerns regarding security and space efficiency.

Threshold systems in classical cryptography have applications in data storage and transmission. They are also used in key dispersal in which the file has been encrypted only once but to obtain the decryption key at least p pieces within a p-k-n system need to be brought together. Such use is different from the one we are advocating in the present paper for instead of different servers or parties or pieces of information, we speak of multiple photons. It is as if each photon is being viewed as a separate channel.

DESIGNING A THRESHOLD SYSTEMS IN QUANTUM CRYPTOGRAPHY

We seek a (p-k-n) threshold quantum cryptography system where if the number of photons being exchanged is less than p, the system is completely secure, when it is between p and k, the system is partially secure, and if it exceeds k, the system is insecure. BB84 is a 1-1-1 system (if $p = k = n$, the threshold property doesn't exist). We are looking at ways to increase the value of p substantially and also leave open the possibility of using redundancy in the transmission so that the communication system can deal with lost photons.

Like BB84, the quantum threshold system must take advantage of the no-cloning theorem. Since the transmitted photons are in an unknown state, the eavesdropper must use tomography to determine the state. Therefore, the number of photons that can be siphoned off by the eavesdropper is less than the minimum required to determine its state. They should also be replaced with other random photons so that the receiver does not know that the medium is tapped.

One system that lends itself to threshold cryptography is the three-stage quantum cryptography protocol (TSQC) [12]. In its basic form it is based on random rotations (or other commutative operators) which can tolerate multiple (up to a certain extent) copies of the photons without jeopardizing security. This protocol can use attenuated pulse lasers rather than single-photon sources in the quantum key exchange, which extends the transmission distance and its use in a variety of implementations [13]-[15]. This basic protocol has been implemented in the laboratory for both free-space and optical fiber [16],[17].

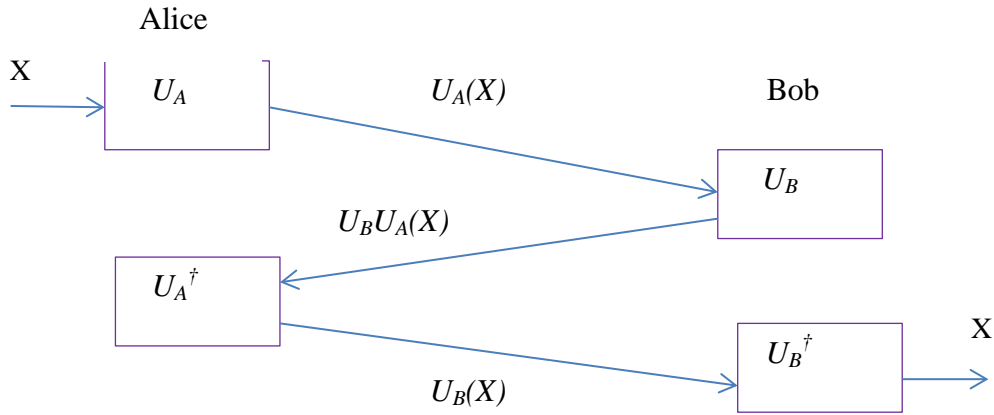

**Figure 1.** The TSQC protocol in which the transformations are random rotations on each bit

Unlike the BB84 protocol which is vulnerable to siphoning of photons in an attenuated pulsed laser system, the TSQC protocol is protected against such an attack since the actual quantum state of the key is never revealed in the communication. This property is of significant importance in the potential use of quantum cryptography in the practical network environment where the optical path is extended beyond trusted routers.

## THE INTENSITY- AND STATE-AWARE TSQC

Intensity- and state-aware versions of the TSQC protocol may be readily devised [18],[19]. In the intensity-aware protocol, the intensity of the beam generated by Alice is known to both the communicating parties. For the sake of simplicity of presentation it will be assumed that there is no loss in intensity during exchange between Alice to Bob. In practice, this loss can be factored in the protocol. The protocol is associated with two parameters: α, the fraction of the incoming beam used by Alice and Bob to determine the beam's intensity, and β, the fraction of the incoming beam diverted by Eve to determine the state of the photons using tomography. The values of α and β are related to light intensity being used.

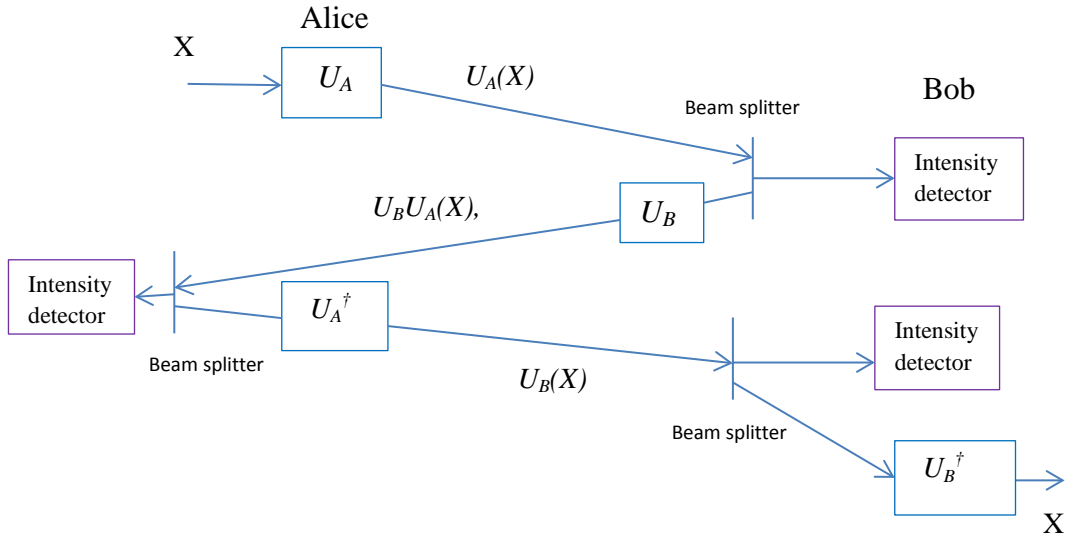

**Figure 2.** Intensity-aware TSQC protocol

The protocol proceeds as follows if the eavesdropper is absent:

*Step 1*. Alice uses photons of intensity I (which is publicly known) to Bob and performs a polarization rotation of angle θ.

*Step 2*. Bob uses a partially silvered mirror (or some suitable beam-splitter) to divert fraction α of the incoming beam to measure its intensity. He applies another random rotation φ on the remaining beam of intensity I(1- α) and transmits it back to Alice.

*Step 3*. Alice receives the beam of intensity I(1- α) and diverts a fraction α of it to measure its expected intensity and undoes her random rotation of angle θ and retransmits the beam of intensity I(1- α)$^2$ to Bob.

*Step 4.* Bob uses the intensity-testing portion of it $I(1-\alpha)^2 \alpha$ for checking it and uses the rest of the beam of intensity $I(1-\alpha)^3$ (by undoing his earlier rotation) to determine the value of the qubit.

When the eavesdropper (Eve) is present, she will be detected by Alice and Bob if the intensity of the beam reaching them in the three stages is less than expected. The constraint on the maximum number of photons being used in the transmissions between Alice and Bob is much less severe than in BB84. If the transmission involves a single photon, or just a few photons, the system is provably secure.

Let us assume that Alice and Bob have agreed to use $s=2^r$ different rotation angles. Eve needs to have access to each of the three transmissions to be able to break the system. Assuming that she needs *m* photons to determine which one of the *s* rotation angles were used by Alice and Bob, the total number she needs to siphon off is at least 3*m*. If the photon source produces 6*m* photons, the siphoning off of 3*m* photons would reduce the intensity by a factor of 2. The intensity used could be a clearly defined variable of the system and any reduction in it would be obvious to Bob and therefore the eavesdropper would be found out.

The variables α and β are in an inverse relationship with respect to the total reduction in intensity.

Let it now be assumed that Alice and Bob will detect any reduction in intensity that is greater than *g* percent. Two cases are possible:

(i) The value of g is constant (and presumably known to Eve) in the communication process.
(ii) The value of g varies randomly (upon mutual agreement by Alice and Bob) and its change provides an additional level of security.

Constant g
Table 1 provides the values of α (Alice's and Bob's diversion probability to measure beam intensity) and β (Eve's siphoning probability) so that the total reduction in intensity is less than 20% (*g*=0.2).

When α is large (say 0.10), Eve cannot do any diversion of 1 percent or higher because that would take the overall intensity reduction over 20%. When α is moderate, say 0.05, she can choose values up to 0.03.

Once the value of g has been chosen, Alice and Bob would like α to be as large as possible so that their use alone brings the received intensity just above the value I(1-g). On the other hand, they wish to use as large a value of I as possible so that the transmission can support a longer distance and that it can also deal with lost photons in transmission. A larger I makes it possible for Eve to use a smaller β.

**Table 1.** α (Alice's and Bob's diversion probability to measure beam intensity) and β (Eve's siphoning probability) for overall intensity reduction that is less than 20% (*g*=0.2)

| α \ β | 0.01 | 0.02 | 0.03 | 0.04 | 0.05 | 0.06 | 0.07 | 0.08 | 0.09 | 0.10 |
|---|---|---|---|---|---|---|---|---|---|---|
| 0.01 | 0.951 | 0.931 | 0.913 | 0.894 | 0.875 | 0.857 | 0.839 | 0.821 | 0.803 | 0.786 |
| 0.02 | 0.922 | 0.903 | 0.885 | 0.867 | 0.849 | 0.831 | 0.814 | 0.796 | | |
| 0.03 | 0.895 | 0.876 | 0.859 | 0.842 | 0.824 | 0.807 | 0.790 | | | |
| 0.04 | 0.867 | 0.850 | 0.833 | 0.816 | 0.798 | | | | | |
| 0.05 | 0.840 | 0.823 | 0.806 | 0.790 | | | | | | |
| 0.06 | 0.814 | 0.798 | | | | | | | | |
| 0.07 | 0.788 | | | | | | | | | |

Varying g
In each session a different value of g is used, or the value of g is changed within a session itself and information on this change is negotiated by the two parties as a part of the protocol.

Note that in a practical system, $m >> s$, therefore, the laser source output can consist of many photons. From information theoretic point of view, the number of photons required to distinguish between *s* rotation angles is $r = \log_2 s$. Therefore, the minimum number of photons that need to be siphoned off by the eavesdropper is $3\log_2 s$. In a practical system, the number of photons to be siphoned off will be significantly greater.

If the technique of using *s* photons each fed into detectors aligned to different polarizations is used then one requires a total of *3s* photons to be siphoned off by Eve to determine the three different polarization angles. If the detectors used by Bob and Alice to check the intensities at intermediate stages can determine reduction of intensity by a factor of one-half, then the source can use *6s* photons in each burst. In reality the constraints are less severe as the bases used by Alice and Bob are arbitrary as long as they are in the same plane.

## SIPHONING ATTACK

Here we consider the question of state decoding in state aware quantum cryptography using a practical method of tomography that can help detect Eve (the eavesdropper) who injects randomly polarized photons to compensate for those she has siphoned off for state detection. Her strategy is to siphon off only as many as she can so that she is not detected and the communication is still possible. This would mean that the randomly polarized photons introduced by her should not swamp the legitimate signal and appear as random channel noise. In other words, Eve can only operate by remaining under the expected noise within the system.

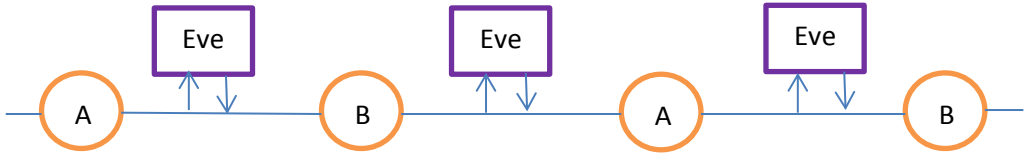

**Figure 3.** Siphoning attack by Eve

Eve can fool Alice and Bob into believing that no photons have been siphoned off if she injects the number of random photons equal to the ones she has taken from the beam. But this can only be done at a cost to her ability to tap the exchange in the subsequent passes. If she withdraws g percent of the photons and injects the same number of some random polarization, the probability that she will be able to determine the polarization angles of the siphoned photons will go down in the next two passes. Furthermore, Bob would know that siphoning had taken place if the noise corresponding to the injected photons exceeds the SNR expectations that he has of the medium.

Let $\alpha$ be the proportion of photons siphoned off by Eve in each pass. After the first pass the proportion of good photons reaching Bob is $(1-\alpha)$, after two passes Alice receives $(1-\alpha)^2$ good photons, and after three passes at the end of the protocol, Bob receives $(1-\alpha)^3$ good photons. This means that the proportion of bad photons (which resemble noise in their action) is $1-(1-\alpha)^3$.

Let the number of photons produced by Alice in first pass be N. We will use the notation {N,0} to mean that N are good photons and there are 0 noise photons.

> At the end of the first siphoning of $\alpha$ percent photons by Eve, the photon vector has changed to $\{N(1-\alpha), \alpha N\}$.
>
> At the end of the second siphoning of $\alpha$ percent photons by Eve, the photon vector has changed to $\{N(1-\alpha)^2, N(2\alpha-\alpha^2)\}$
>
> At the end of the third siphoning of $\alpha$ percent photons by Eve, the photon vector has changed to $\{N(1-\alpha)^3, N(3\alpha-3\alpha^2+\alpha^3)\}$

The signal to noise ratio (SNR) for Bob after the photons corresponding to X have been received is:

$$SNR = \frac{(1-\alpha)^3}{1-(1-\alpha)^3} \qquad (1)$$

It may be easily shown that the value of SNR remains greater than 1 as long as $\alpha < 0.2062$. For effective working, SNR should be much larger than 10 or so. Figure 4 graphs the relationship between SNR and $\alpha$. In reality, the value of SNR should be much greater than 1 for tomography to be effective.

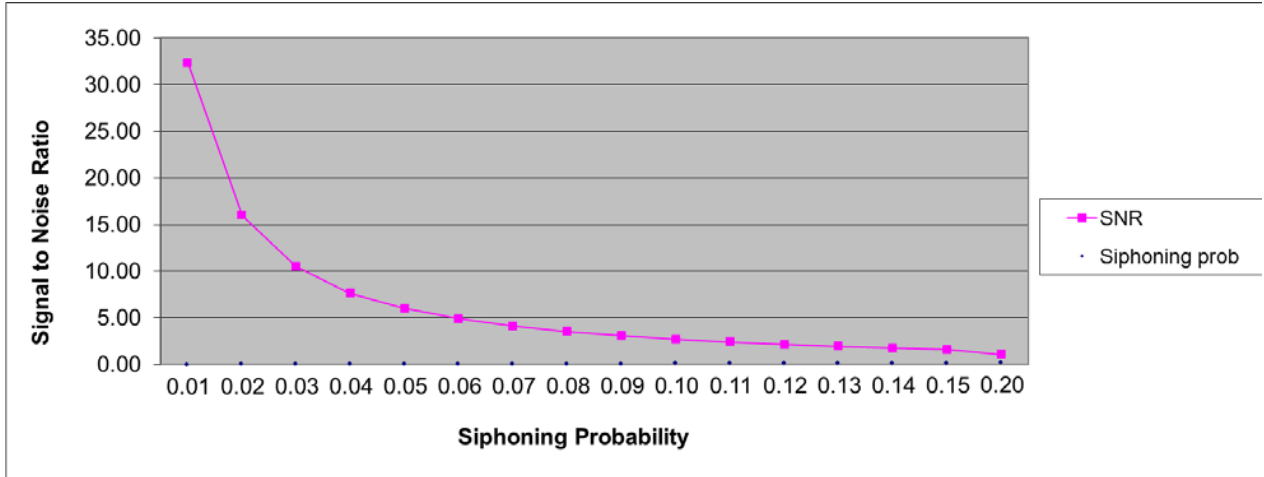

**Figure 4**. SNR with respect to siphoning probability

Notice that as Eve introduces random photons into the stream to compensate for the ones she has siphoned off, the SNR gets degraded both for her as well as for Alice and Bob. Since she needs to siphon off as few photons as she can, she will require more photons in the second and third passes to have a minimum number of good photons.

To see this, consider the example of 100 photons being sent by Alice to Bob in the first pass. If Eve siphons off 20 photons (20% of the total) to ensure that she is able to determine the polarization used by Alice, what Bob receives are 80 good photons and 20 bad ones (the randomly polarized photons put in by Eve. In the second pass, Eve still needs 20 photons for tomography but it is now 25% of the 80 good ones, so the total number of photons she must take out (and replace) is 25. Thus at the end of the second pass, the photon stream has 60 good photons and 40 bad photons. In the third pass, Eve need to siphon off just over 33% (or say 34%) of the photons to ensure that she has 20 good photons, but she also has 14 bad photons. Her own SNR in passes 2 and 3 has been 20/5=4 and 20/14=1.43, respectively. The best she or Bob can do is to attain SNR value of 1.5 but she will need to siphon off the entire stream (leaving nothing for Bob) to attain this value.

In the more general case, where Eve siphons off $\alpha_1$, $\alpha_2$, and $\alpha_3$ fractions of the incoming photons, the SNR at the last stage will be:

$$SNR = \frac{(1-\alpha_1)(1-\alpha_2)(1-\alpha_3)}{1-(1-\alpha_1)(1-\alpha_2)(1-\alpha_3)} \qquad (2)$$

Assume that the least number of photons needed by Eve to determine their polarization in the last state is p. Then the total number of photons she needs on the three passes is roughly 3p since the SNR in the earlier stages would be higher and she might be able to do with fewer number of photons.

Since Eve needs to extract at least 3p bits, the bits required by Bob may be taken to be 4p for proper detection. The real number will be much less as Bob does not need to perform state tomography. But if we take this number, then the three-stage protocol is a (p-4p-n) system.

## CONCLUSIONS

This paper argues for the development of threshold quantum cryptography protocols in which the system is secure so long as the number of photons being exchanged between Alice and Bob is below a specified threshold. The BB84 protocol is (1-1-1) whereas the three-stage protocol appears to be (p-4p-n), where p is the least number of photons necessary to determine the polarization state of identically prepared photons.

Some proposed quantum cryptography systems use entangled photons [20], but the technology has not reached a level of maturity so that they can be used in practical schemes [21].

It is hoped that this paper will encourage researchers to look for other systems of threshold quantum cryptography.

*Acknowledgements.* This research was supported in part by the National Science Foundation grant CNS-1117068.